# Fault-Tolerant Performance Enhancement of DC-DC Converters with High-Speed Fault Clearing-unit based Redundant Power Switch Configurations

Tohid Rahimi
*Faculty of Electrical and Compuer Engineering*
*University of Tabriz*
Tabriz, Iran
rahimitohid@yahoo.com

Hossein Khoun Jahan
*Faculty of Electrical and Compuer Engineering*
*University of Tabriz*
Tabriz, Iran
hosseinkhounjahan@yahoo.com

Armin Abadifard
*Faculty of Electrical and Compuer Engineering*
*University of Tabriz*
Tabriz, Iran
a.abadifard93@ms.tabrizu.ac.ir

Mohsen Akbari
Faculty of Electrical Engineering
K.N. Toosi University of Technology
Tehran, Iran
mohsen.akbari.eng@gmail.com

Pedram Ghavidel
*Faculty of Electrical and Compuer Engineering*
*University of Tabriz*
Tabriz, Iran
p.ghavidel93@ms.tabrizu.ac.ir

Masoud Farhadi
*Department of Electrical and Computer Engineering*
University of Texas at Dallas
Dallas, Texas, USA
masoud.farhadi@utdallas.edu

Seyed Hossein Hosseini
*Department of Electrical and Computer Engineerin*
University of Tabriz
Tabriz, Iran
Hosseini@tabrizu.ir

*Abstract* — **Fault detection and reconfiguration in fault-tolerant converters may complicated and necessitate using all-purpose microprocessors and high-speed sensors to guarantee the satisfactory performance of power converters. Therefore, providing fault-clearing feature without increasing the processing and sensing burdens and reducing the transition time between faulty to normal state are of great importance. This research proposes a new redundant-switch configuration to address the mentioned challenges. The proposed configuration uses one diode and two fuses to eliminate the faulty switch and replace the reserve one, spontaneously. Open-circuit fault in the proposed configuration is clarified instantly. Moreover, the short-circuit fault is dealt as an open-circuit fault by using a fuse. Thus, the fault-tolerant feature of the proposed switch configuration is achieved without using a complex, versatile and multifaceted fault managing unit. Resultant behaviors of the case studies are derived using MATLAB/SIMULINK. Also, steady-state thermal distribution of power switches which are implemented on a monolith heat sink, analyzed in COMSOL Multi-physics environment. Finally, the viability of the proposed configuration is demonstrated by a laboratory-scaled prototype.**

***Keywords—Fault-tolerant converters, power switch failures, self-reconfigurable feature, redundant-switch configurations.***

## I. Introduction

As the power converter market is growing rapidly, the reliability of the systems gain more and more importance, especially in critical and sensitive applications [1-4]. From reliability point of view, the power semiconductor switches are one of the most vulnerable elements [5], [6], [18]. Identifying influential factors on the reliability level should be considered to evaluate and enhance the reliability of the converters [8]-[10]. Therefore, the designers' aim is to improve the reliability of power converters by using active and passive methods. In other words, the reliability evaluation and improvement of power converters are two basic duties of reliability engineers. To improve the reliability of power converters, many fault detection and reconfiguration methods are discussed in literature; all the presented methods require redundant switches, modules, or even reserve systems to guarantee the satisfactory performance of power converters. To tackle the problem, this paper aims to propose a redundant-switch configuration that spontaneously isolates the faulty part and replaces the reserve switch. Therefore, it eliminates the need for any hardware and/or software required to detect and isolates the faulty section.

Fault detection is an important step in fault-tolerant converters. Open- and short-circuit faults of power switches are two possible faults that may cause the entire power converter to fail. In [7], a low-frequency sampling principle is used to detect open-circuit switch fault of two-level and natural point clamped (NPC) inverter. This method depends on the system model and threshold values. In [8], fault diagnostic techniques have been proposed to detect open- or short-circuit faults in the conventional boost converter. The mentioned techniques consist of the primary and secondary algorithms. After fault detection, the redundant switch is replaced via a bidirectional switch that increases power loss in the post-fault condition. In [9], a high-speed fault detection method for the conventional boost converter is presented that increases the processing burden of the processor and requires high-speed sensors. In [10], a fault-tolerant phase-shifted full-bridge converter has been presented, which consists of fault detection and output voltage compensation configurations. It requires additional hardware tools for post-fault operation. For a cascaded Quasi-Z Source converter, a fault-tolerant scheme has been introduced in [11] that isolate the faulty module. However, many numbers of relays have been used in the hardware of the fault management system. A complex flowchart increases the processing burden and necessitates using powerful and multifaceted microprocessors such as Digital Signal Processor (DSP), thus using many relays are the main disadvantages of the proposed converter in [12]. In [13], a new reconfiguration pattern and switching strategy are proposed to manage faulty phases. However, due to using a conventional microprocessor, the reconfiguration strategy has low speed. A fault diagnostic method for synchronous boost converter based on a predictive current emulator model is presented in [14]. In this method, the sampling frequency must be N times the switching frequency, which necessitates high-speed sensors and microprocessors. The switch and diode conditions in the conventional buck converter are monitored by diode voltage sensing. This method is simple, but the method depends on threshold values. In summary, the open- or short-circuit fault detection methods that are presented in previous papers require to watch out voltage and/or current of semiconductor devices [15].

The fault detection strategies in [8]-[10], and [13] are based on the sign of the inductor current slope, which require processing the switching pattern, hence, the processing burden is high and high-speed precise-current sensors are needed. The other problem comes with the switching frequency, with increasing this frequency, spotting the sign of the inductor current slope would be very difficult. Furthermore, the need for monitoring the current of the faulty switch causes extra complexity in the existing reconfiguration strategies. Also, some of the mentioned strategies have limited applications. For instance, the strategies in [8], [9] can only be applied to single-switch converters. What is more, the transition time from faulty to the post-fault configurations is high in some of the mentioned works [11], [12].

In order to simplify the fault clearing mechanism and address the above mentioned challenges, this paper proposes a new redundant-switch configuration that does not need any high speed open- and short-circuit fault detection systems. In this configuration the instantaneous fault detection and isolation are achieved through two fuses and one diode. Therefore, by employing the proposed configuration, the complex stage to detect and isolate the fault is eliminated, the processing burden is reduced and the need for high-speed and precise sensors are eradicated. Hence, it would be possible to implement a simple and cheap fault-managing unit by utilizing some low-speed and cheap sensors along with a simple microcontroller.

The remainder of this research is organized as follows. In the next section, the three proposed switch-configurations for DC-DC converters are discussed. The simulation waveforms are shown in section IV to validate the proposed fault-tolerant converter capability. Also, Finite element (FEM) analyses show the steady-state temperature distribution of power switches. A discussion on technical point of the proposed switches is presented in Section VI. Transition state from healthy to faulty conditions is validated using experimental results in section VII. Finally, the overall work is concluded in section VIII.

## II. The Proposed Switch Configuration

In this paper, a new redundant-switch configuration proposed that eliminates the need for a fast open-circuit fault detection and managing unit. In this configuration the short-circuit fault of the switch can be dealt as an open-circuit fault by connecting a fuse as series with the switch. The proposed redundant-switch configuration, which is shown in Fig. 1 (a), consists of two switches and one diode. As shown in Fig 1(a), the switch $S_1$ is placed in the main branch, and the $S_2$, along with the diode $D$ is placed in the reserve branch. In addition, the fuses are in series with power switches to isolate the faulty branch in short-circuiting faults.

The switching order of the proposed configuration is shown in Fig.1 (b). The $S_1$ must be turned on firstly, and after that, the switch $S_2$ is turned on. This strategy keeps the diode in reverse bias mode. Therefore, $t_d$ should be large enough to make switch $S_2$ act like an off switch when switch $S_1$ is in transient switching condition. $t_d$ must be longer than $t_{on}$ of the power switch to ensure the total current does not flow through the main branch.

It is acknowledged that power converters experience transient modes in dynamic change states. In order to prevent misdoing of the switch during dynamic change states, the fuses ($F_1$ and $F_2$), utilized in the proposed configuration, and should have a specific time-delay (Fig. 1(c)). Additionally, semiconductors switches typically can tolerate short-circuit current for a specific duration to allow the controller to turn them off before causing any damaging. The fuses must isolate the switch before dangerous conditions come about.

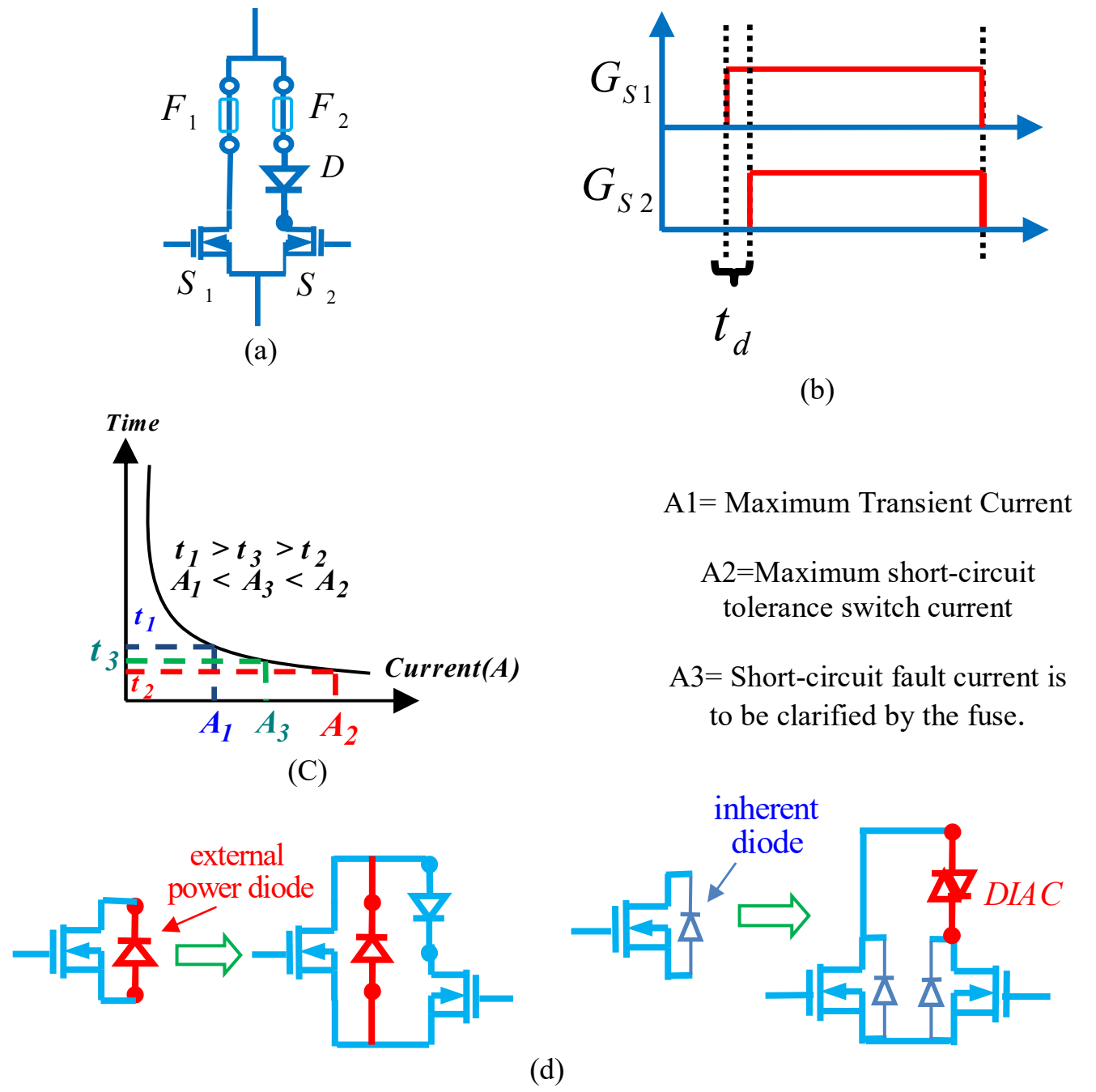

Fig. 1. Proposed redundant-switch configuration, (a) proposed redundant-switch configuration exemplified by MOSFET switches, (b) switching pattern (c) fuse requirements, and (d) proposed configuration for ac applications.

The proposed switch configuration is not a blind response to a fault event. After spontaneous fault clearing by the proposed redundant-switch configuration, the fault managing system can detect the fault and recognize the new conditions through the simple and cheap sensors. The proposed switch configuration can be modified for AC application as seen in Fig. 1 (d).

When the $S_1$ is shorted-circuited, the branch current increases significantly. If the fuse opens the faulty branch, the branch current is going to increase, and the reserve switch dose not conduct the branch current. On the other hand, the gate signal of the reserve switch is modulated by the control system to make its width as narrow as possible. Thus, as the fuse has opened the faulty branch, the reserve switch would face the short-circuit current for very small duration. Under this condition the fault managing unit is not required to detect the short-circuit fault quickly to keep ready the reserve switch in standby mode to replace the faulty switch. The sensor in this method would have adequate time to detect the short-circuit fault. Thus, there is no need for the sensors to be fast and precise to detect the short-circuit fault.

When the $S_1$ is open-circuited, the reserve switch replaces spontaneously without requiring any sensors and fault-detection strategies. However, to make the fault-managing unit aware of the new condition, another ordinary sensor should be used to detect the current condition of the reserve branch.

## III. Proposed Fault-Tolerant Boost Converter

During last decade, different types of DC-DC converters have been proposed in the literature [19]-[22]. However, most of them use the same philosophy to generate dc voltage. In this section, the application of the proposed configuration in a conventional boost converter is assessed (Fig. 2 (a)). Switching signal generation under normal conditions is shown in Fig. 2 (b). The control unit consists of two proportional integral (PI) controllers and two sensors. First PI controller is employed to regulate the output voltage. The output of the voltage compensator is the inductor current reference. The second controller is used to the control this current and keeps the current under the maximum value. As discussed in previous sections, the open-circuit is cleared instantly. As soon as the open-circuit fault occurs, the reserve switch conducts whole the branch current. So, to ensure service continuity of the single switch converter after the open-circuit fault occurrence, remedial actions are not necessary. Thus, when the open-circuit fault occurs, the fault can be cleared rapidly and the desirable output voltage is guaranteed reliably. This advantage is an interesting merit of the proposed redundant-switch configuration in single-switch dc-dc converters.

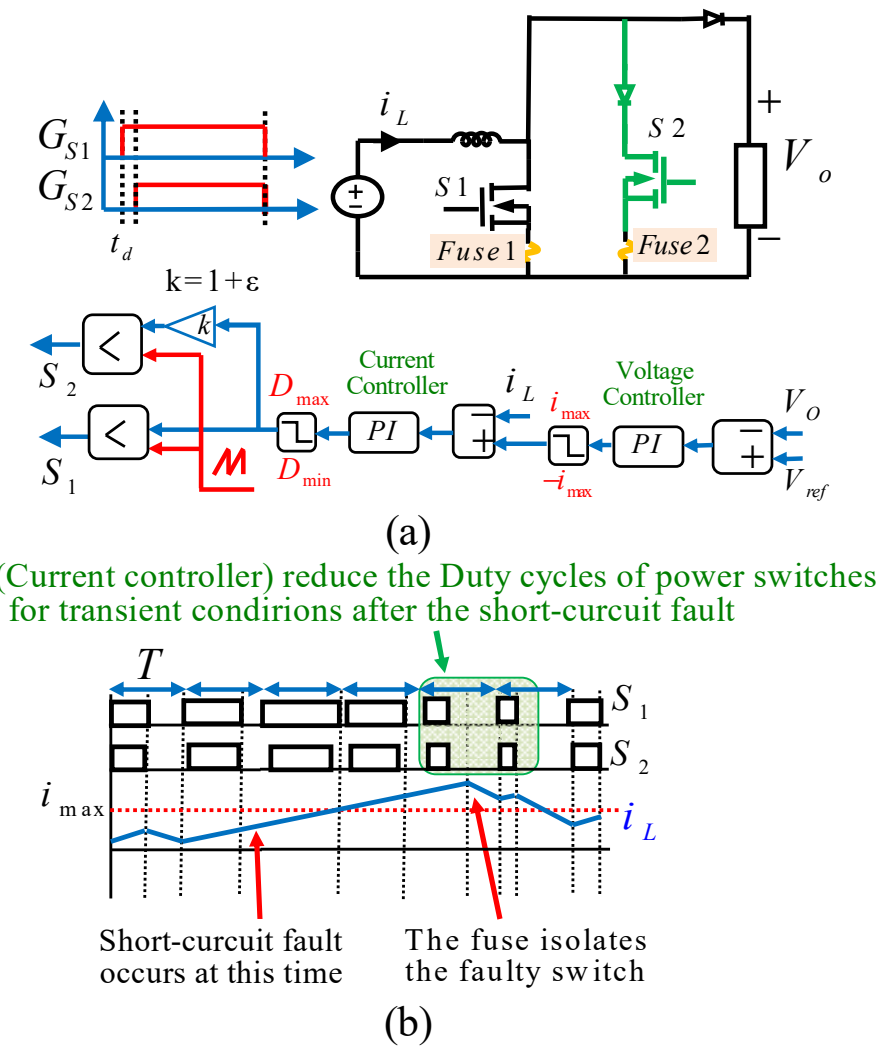

Fig. 2. Single-switch boost converter with proposed redundant-switch configuration, and (b) switching pattern reforming duration after short-circuit fault and faulty switch isolation conditions.

TABLE I. Comparison of Fault-Tolerant Strategies

| References | | [8] | [9] | [11] | [12] | [13] | [15] | The Converter of Fig. 2 |
|---|---|---|---|---|---|---|---|---|
| Detection parameter | | Inductor current slope | Inductor current slope | Relay based | Relay based | Inductor current | DC-link current derivative sign | Inductor current slope |
| Detection and Clearing Fault time | OP | > 4T$_{sw}$ | < 1T$_{sw}$ | 20ms | 5ms | < 1T$_{sw}$ | > 3T$_{sw}$ | << 1T$_{sw}$ |
| | SC | Fuse-based | Fuse-based | 20ms | 5ms | Fuse-based | null | Fuse-based |
| Type of fault | | OP, SC | OP, SC | OP, SC | OP, SC | OP, SC | OP | OP, SC |
| Reconfiguration capability discussion | | Yes | Yes | Yes | Yes | No | No | Yes |
| High speed sensors and process are required? | | Yes | Yes | No | No | Yes | Yes | No |
| Fault-tolerant capability can be achieved by cheap sensors and microprocessors? | | No | No | Yes | Yes | No | No | Yes |
| Observing faulty switch current for reconfiguration actions is required? | | Yes | Yes | Yes | Yes | No | No | No |
| The reserve switch must stand a large current after short-circuit fault? | | No | Yes | Yes | Yes | null | null | No |
| Number of additional required elements for providing fault-tolerant capability? | | 1 Triac 1 Switch 2 Fuse | 1 Switch 2 Fuse | 9 Relays For three modules | 1 Switch 2 Relays For each Phase | null | null | 1 Diode 1 Switch 2Fuse |

When the short-circuit fault of the main power switch occurs, this fault is cleared by the fuse and the inductor current value is high. If the reconfiguration order is issued after isolation of the faulty switch, the reserve switch may be damaged due to conducting the high current for transient states.

Thus, an effective measure should be taken to tackle this problem. In many papers such as [8], [9], [11] and [12], this challenge is not notified. Along with the proposed redundant-switch configuration an improved control strategy is put forth in the present letter to alleviate the mentioned problem, the suggested control strategy is exhibited in Fig.2 (b). According to this figure, to deal with the short-circuit fault, a suitable fuse is employed. The characteristic of the suitable fuse is discussed in Fig.1 (c).

To tolerant the short-circuit fault, the suggested control system in the converter can improve the capability of the converter in managing faulty conditions. When short-circuit fault occurs, the inductor current rises linearly, and it tends to go beyond the maximum allowed value. At this moment, the current compensator reduces the duty cycle values and makes it narrower. Therefore, when the fuse in the faulty branch isolates the faulty switch, the reserve switch withstands the high current value for a short interval (during the narrowed duty cycle), thus, it passes the transient state safely without facing the high-current in a long interval. This advantage is obtained because the current compensator reduced the duty cycle effectively, as shown in Fig.2 (b).

Since in the suggested control loop, only monitoring of the average value of current of the inductor is sufficient, the sensors do not need to be fast and precise. This can simplify the sensing procedure and reduce the cost of the system, compared to the methods that are presented in [8], [9], [13] and [15]. The fault tolerant methods in these references require high-speed and precise sensors and have a huge processing burden. To highlight the merit of combination of the proposed redundant-switch configuration and improved control strategy, a comparison of the proposed configuration with is the state-of-the-art methods is worked out and the results are provided in Table I. According to these comparisons, it can be noted that the proposed configuration is one of the best in terms of fault clearance in both open-circuit and short-circuit cases. In this table, "OP" and "SC" are indicators of open-and open-circuit faults, accordingly.

## IV. Simulation Results

For simulation, a single-phase boost converter with specific characteristics is considered. The input voltage, the output voltage, the output power and the inductor values are 50 V, 200 volt, 200 Watt and 1.8 mH, respectively. In this section, simulation results of the proposed topology shown in Fig. 2(a) and suggested in Ref. [9] are presented to show the importance of limiting duty cycle of the reserve switches for post-fault conditions. In this regard, the main switch fails with the short-circuit fault at $t$=0.25(s). Fig. 3 shows the behavior of the converter equipped with the proposed switch configuration and closed loop control which is discussed in the previous section.

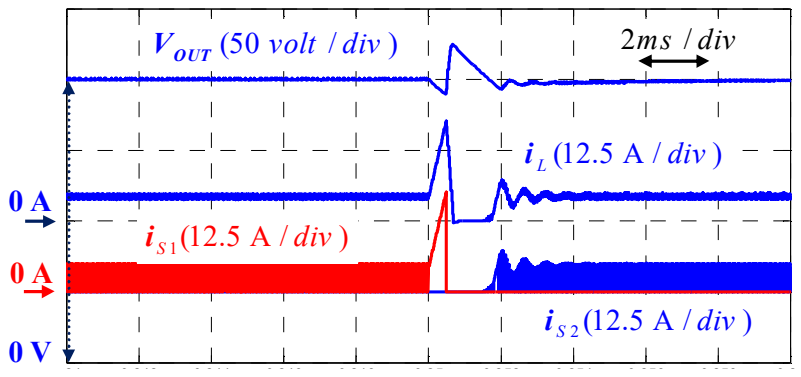

Fig. 3. Simulation results of the boost converter with proposed redundant-switch configuration and the closed-loop control method.

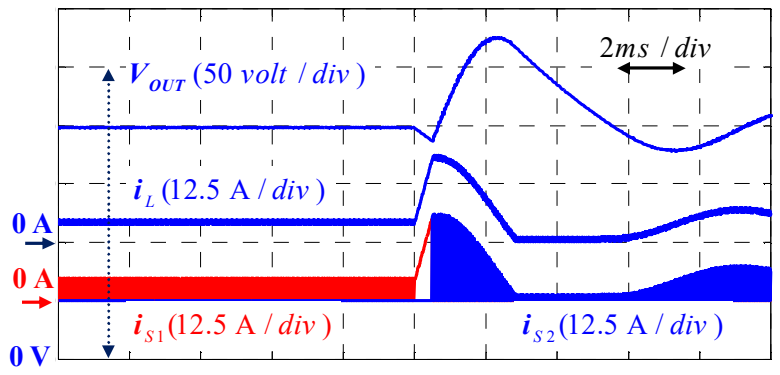

Fig. 4. Simulation results of the boost with the employed method in [9].

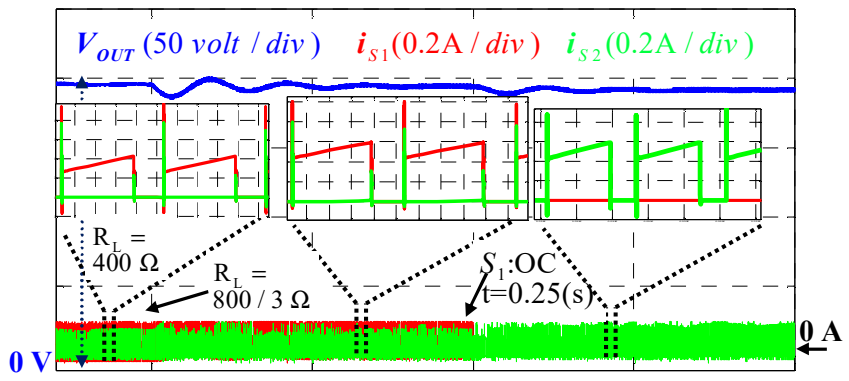

Fig. 5. Validation of the proposed configuration by applying step change.

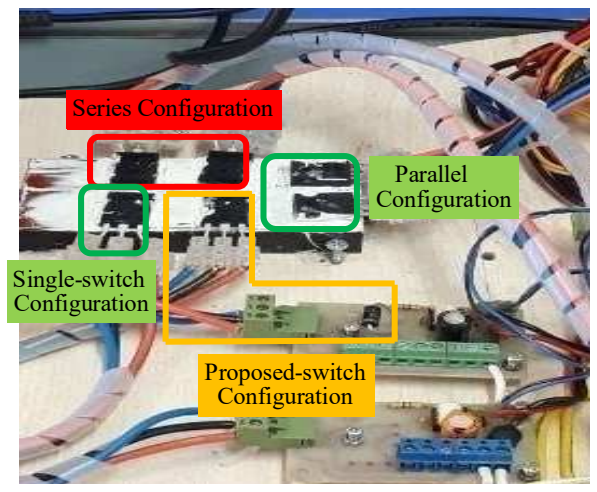

Fig. 6. The Power swithes layout on the surface of the heat sink.

As seen in this figure, the reserve switch has not conduct short-circuit current after isolation of the faulty switch and the converter can keep its normal operation after clearing the fault. The reserve switch in the topology of Ref. [9] has to conduct the short-circuit current after clearing faults which decrees the lifetime of the reserve switch (Fig. 4). Limitation the duty cycle of the reserve switch for transition state from healthy condition to the faulty condition reduces the transient time and the converter can continue its normal operation quickly.

It is worth noting that the proposed structure in the dynamic change state of the converter still has the expected behavior that is shown Fig.5. In this section, a step change test is done to see whether the proposed configuration can be immune to dynamic operations. Since no control feedback was taken to replace the healthy switch instead of the faulty switch, the reserve switch in the proposed configurations (fig. 1(a) and fig. 1(d)) is kept in non-conducted mode, automatically. To derive Fig.5, detail model of the power switches and diodes are considered. The detail characteristics of the used power switch and diode can be extracted from their datasheets.

## V. THERMAL ANALYSE

To perform a thermal analysis for comparison, the COMSOL Multiphysics, which is a finite element (FE) method-based simulation platform, is employed. In this work, as it can be seen in Fig. 7, four configurations are considered for power switches of the single-phase boost converter. The switches all are mounted onto the same heat sink with the natural air cooling. To simulate the real heat sink effect, a method introduced in [16] and [17] is utilized. Temperature distributions obtained from FE simulations are shown in Fig. 7. It can be seen that the temperature in the two-switch operation with the parallel or series configuration is higher compared to the single–switch operation. Moreover, it is realized that in the series configuration, the heat sink base would be cooler due to the lower power loss (heat) created within the switches in comparison to the parallel configuration.

## VI. DISCUSSION

Due to power losses of the additional diode (or DIAC), the efficiency of the proposed configuration is low in post-fault conditions. However, the proposed configuration does not need any open circuit fault detection in the first step (Hence, a fuse is replaced with fault detection and isolation unit that can reduce the complexity and cost of the converter). Therefore, a simple processor and so a low cost fault detection system is required, and the probability of the successful operation of the fault detection system is high.

As discussed in previous literature, to implement fault-tolerant converters to continue their normal operation, extra components or power stages are necessary. These extra units increase conduction and/or switching losses [8, 10]. To select the added diode or DIAC for the proposed configurations, the voltage across diode must be lower than the feed-forward voltage of the diode. Thus, we have:

$$V_{S1|V_{gs}=1} - V_{S2|V_{gs}=1} < V_{Feedforward} \quad (1)$$

As discussed here, increasing the conduction losses due to added diode in post-fault condition is not a fatal drawback of the proposed configuration in comparison with other fault-tolerant structures. Despite all the reasons discussed, in this section a modified version of the proposed switch configuration is proposed that is seen in Fig. 8. After clearing fault, using cheap sensors or methods, operation of S2 instead of S1 can be detected and the relay would be closed after fault detection. Thus, power losses problem of diode can be solved.

## VII. EXPERIMENTAL RESULTS

In this section, experimental results are provided to verify the performance of the proposed redundant-switch configuration. A (300-W, 200V) three-phase interleaved converter, as a case study, is implemented; this prototype is exhibited in Fig. 9(a). One phase of this conventional three-phase converter is equipped with the proposed switch configuration. Other phases are equipped with series and parallel switch configuration. According to Fig.9 (b), the parallel power switches carry half of the phase current. So, each power switch conducts a certain portion of the branch current. With the simultaneous operation of two switches, each switch withstands half of the nominal voltage, as seen in Fig. 9(b). As seen in Fig. 9 (d), in the proposed redundant-switch configuration, the reserve switch does not conduct when the main switch carries the whole phase current. Therefore, the auxiliary switch $S_2$ stays in reserve mode (Fig. 9(d)).

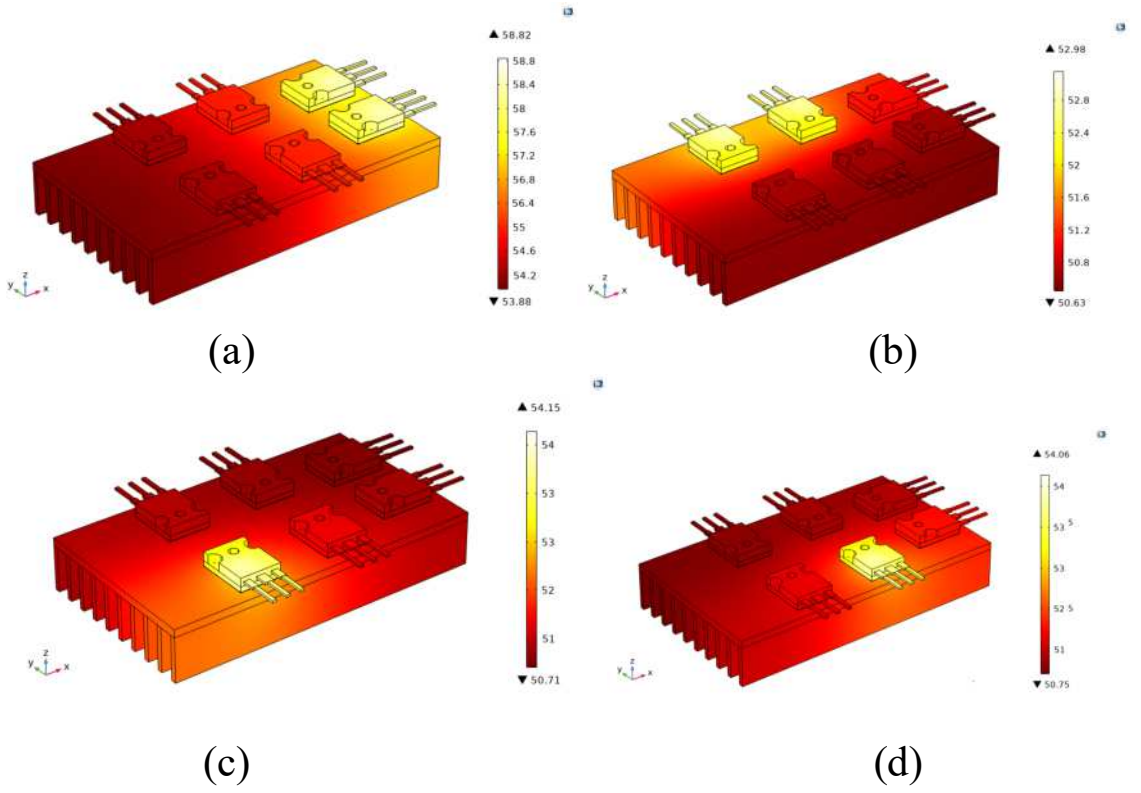

Fig. 7. Top surface temperature distribution of conditions for operation of single-phase boost converter and considering Ta=25 ℃, Po=100 W, and Vo=200 V equipped with: (a) Parallel configuration, (b) Series configuration, (c) Single switch.

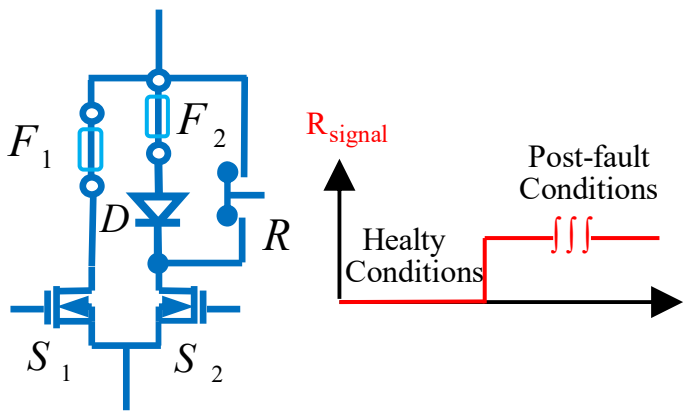

Fig. 8. The third proposed redundant-switch configuration.

In the experimental test, the open-circuit fault occurs when the main switch ($S_1$) in the proposed configuration is on. The reserve switch conducts immediately without any additional detection and reconfiguration action. Fig. 9(e) shows the results for the open-circuit fault. In the following, the results are shown in Figs. 9 (f), and (g) for the short-circuit fault.

In this case, when the main switch ($S_1$) is on, a short-circuit fault is applied to this switch. When the fuse ($F_1$) isolates the faulty switch, the reserve switch ($S_2$) conducts the branch current spontaneously. As shown in the provided results, the faulty switch can be isolated, and the reserve switch can be replaced without using high-speed sensors and microprocessors.

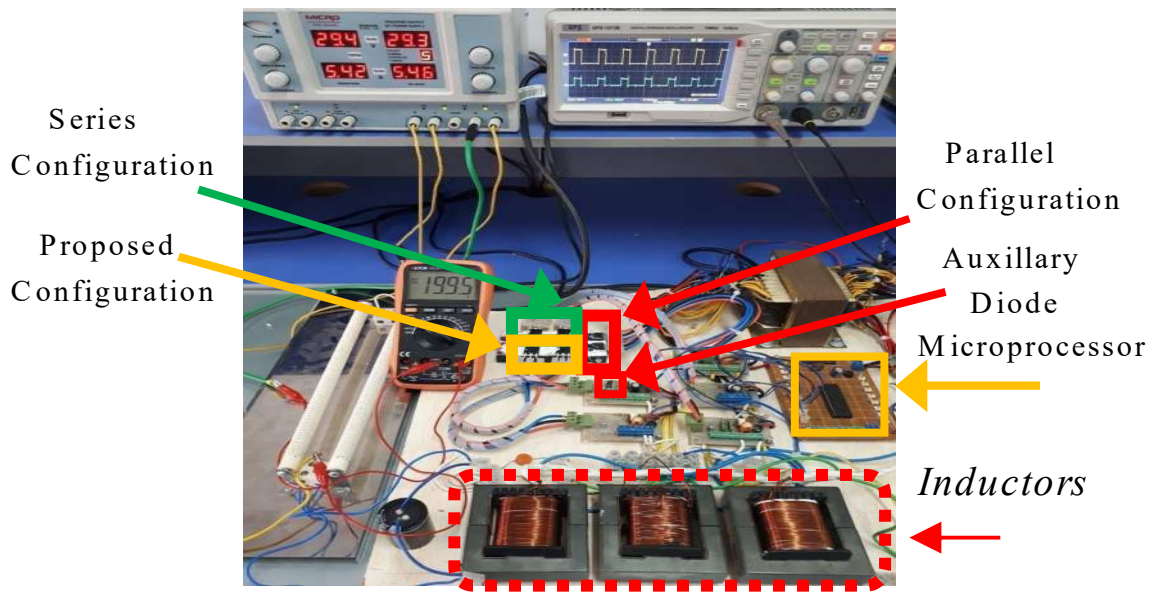

(a)

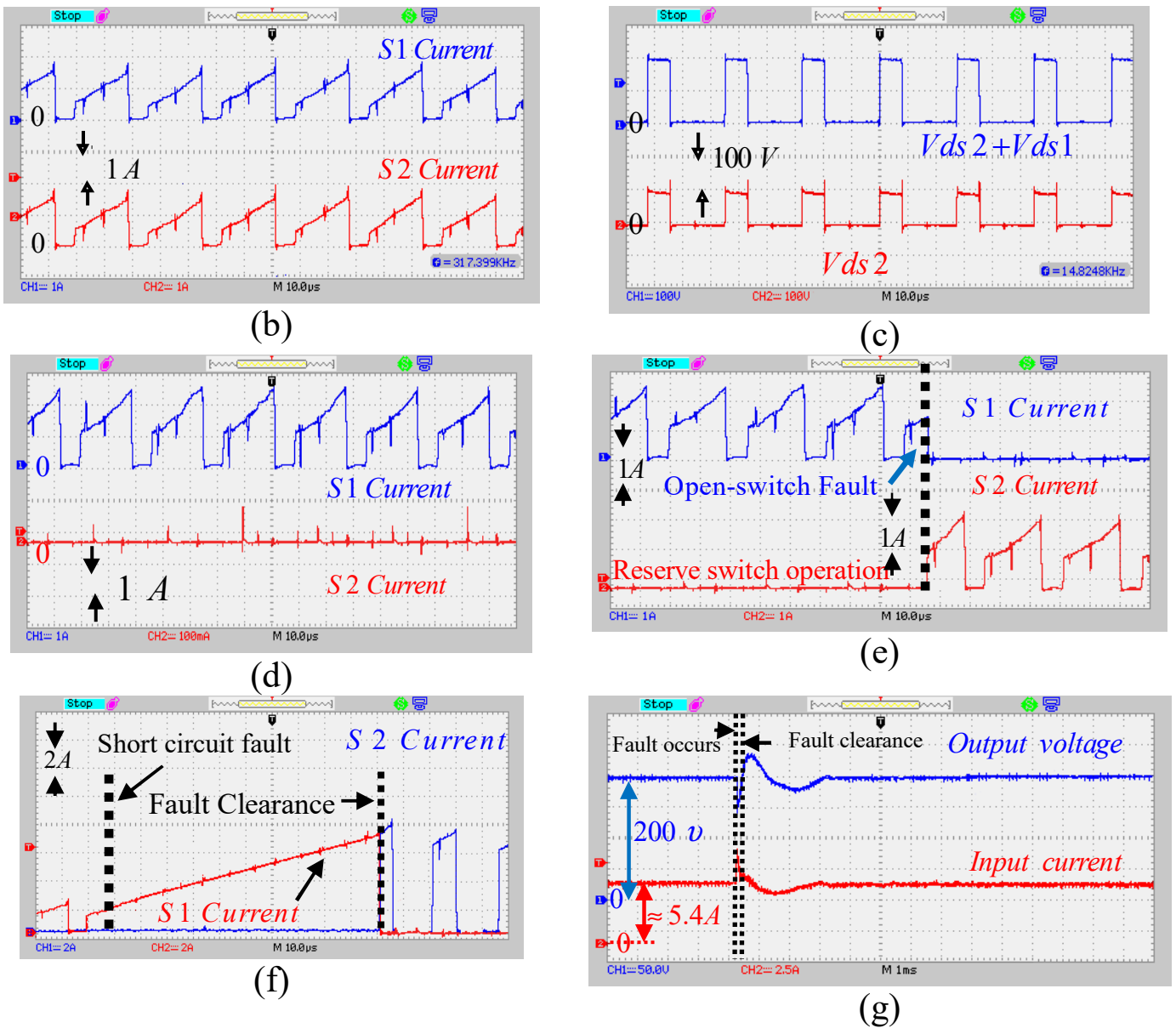

Fig. 9. Validation of power converter reliability equipped with proposed redundant-switch configuration (a) Experimental setup; (b) Switches currents in parallel configuration, (c) voltage across S2 and total voltage in series configuration, (d) switches currents of the proposed configuration, (e) Open circuit fault detection and isolation by the proposed redundant-switch configuration, (f) short circuit fault detection and isolation by a fuse and automatic current conducting by reserve switch, and (g)output voltage and input current in post-fault condition.

## VIII. Conclusion

In this research, new redundant-switch configurations were proposed for single-switch converters. The proposed configurations can clear open-circuit instantaneously. Also, the Short-circuit fault is cleared by the fuse and then the reserve switch operates instead of the faulty power switch rapidly. Thus proposed configurations can, reduce the processing burden and provide a fast and spontaneous fault clearance capability by using ordinary sensors. The performance of the conventional boost converter with the proposed switch configuration are analyzed and its control section is adapted to reduce the duty cycle of the remained switch for transient conditions after the main switch fault. Employing this strategy can reduce the high current stress of the reserve switch after short-circuit fault clearances effectively that is considered in the previous literature rarely.